\documentclass[aps,prl,preprintnumbers,reprint,nofootinbib,showpacs]{revtex4-1}
\pdfoutput=1

\usepackage{graphicx}
\usepackage{epsfig}
\usepackage{amssymb}

\usepackage[mathletters]{ucs}
\usepackage[utf8x]{inputenc}

\newcommand{\TT}{T_{\perp}}
\newcommand{\tT}{\tau_{\perp}}

\begin{document}

\title{Next-to-next-to-leading logarithmic resummation for transverse thrust}
\author{Thomas Becher}
\author{Xavier \surname{Garcia i Tormo}}
\author{Jan Piclum}
\affiliation{Albert Einstein Center for Fundamental Physics. Institut f\"ur Theoretische Physik, Universit\"at Bern,
  Sidlerstrasse 5, CH-3012 Bern, Switzerland}

\date{\today}


\begin{abstract}
We obtain a prediction for the hadron-collider event-shape variable transverse thrust in which the terms enhanced in the dijet limit are resummed to next-to-next-to-leading logarithmic accuracy. Our method exploits universality properties made manifest in the factorized expression for the cross section and only requires one-loop calculations. The necessary two-loop ingredients are extracted using known results and existing numerical codes. Our technique is general and applicable to other observables as well.
\end{abstract}

\maketitle

Event-shape variables are an important tool to characterize QCD effects at colliders. They are designed to measure geometrical properties of energy flow in collisions and provide information about the distribution of particles in the final state. Due to their inclusive nature, they can be computed perturbatively and have only mild sensitivity to hadronisation effects. Such observables were among the first proposed to test QCD and can also help in discriminating new-physics effects against the Standard-Model background. Traditionally event shapes have been mostly used in leptonic collisions, but they are also of great interest in the richer environment of hadronic collisions. 
There is, for instance, a lot of recent work using event shapes as a tool to study jet substructure, and they can also be instrumental in improving our knowledge of some poorly understood aspects of hadronic collisions, such as underlying-event (UE) effects. In this Letter, we provide results for the archetypical hadron-collider event-shape variable, transverse thrust, at an unprecedented level of accuracy.  

A large class of dijet event shapes for hadronic collisions was defined in Refs.~\cite{Banfi:2004nk,Banfi:2010xy}, using only momentum components $\vec{p}_\perp$ transverse to the beam direction, in order to reduce sensitivity to the beam remnants. We will denote a generic transverse event-shape variable by $e_{\perp}$. The classic example is $e_{\perp}=\tT=1-\TT$, where the transverse thrust $T_{\perp}$ is defined as
\begin{equation}\label{eq:def_trans_thrust}
\TT:=\max_{\vec{n}_{\perp}}\frac{\sum_m\left|\vec{p}_{m\perp}\cdot\vec{n}_{\perp}\right|}{\sum_m |\vec{p}_{m\perp}|}\,.
\end{equation}
The sums run over final-state particles $m$. Transverse thrust has been measured at the LHC~\cite{Khachatryan:2011dx,Aad:2012np,Aad:2012fza,Chatrchyan:2013tna,Khachatryan:2014ika} and previously also at the Tevatron~\cite{Aaltonen:2011et}. In the dijet limit $e_{\perp}\to 0$, higher-order terms enhanced by logarithms of $e_{\perp}$ need to be resummed in order to obtain reliable theoretical predictions. This resummation was performed at next-to-leading logarithmic (NLL) accuracy in Refs.~\cite{Banfi:2004nk,Banfi:2010xy} within an automated framework~\cite{Banfi:2004yd} (for leptonic collisions this was recently extended to next-to-next-to-leading logarithmic (N$^2$LL) accuracy in Ref.~\cite{Banfi:2014sua}).

In Ref.~\cite{Becher:2015gsa} we performed an analysis of transverse thrust within the framework of Soft Collinear Effective Theory (SCET) \cite{Bauer:2000yr,Bauer:2001yt,Beneke:2002ph} (see Ref.~\cite{Becher:2014oda} for a review) and obtained a factorized expression for the cross section that permits resummation of terms enhanced in the dijet limit to arbitrary accuracy. For a generic $e_{\perp}$ the factorization formula can be written as ($\otimes$ denotes a convolution)
\begin{equation}\label{eq:fac_cs}
\frac{dσ}{de_{\perp}} =  \sum_{a,b,i,j}\,P^{ab\to ij}_{IJ}\otimes S^{ab\to ij}_{JI} \otimes J_{i} \otimes J_j \otimes B_a \otimes B_b\,,
\end{equation}
where the sum runs over different partonic channels. Here and below, the letters $a$ and $b$ denote initial-state partons and $i$ and $j$ final-state ones. In the above equation, the factor $P^{ab\to ij}_{IJ}$ encodes effects at the hard scattering scale $Q$. It includes two parts: a hard function $H^{ab\to ij}_{IJ}$ and a so-called collinear-anomaly term, which involves hard-scale effects related to large rapidity differences among emitted particles. $S^{ab\to ij}_{JI}$ is the soft function, encoding effects of lower-energy soft radiation; it is contracted with the hard function via the  color indices $I$ and $J$. The jet and beam functions $J_i$ and $B_a$ encode the collinear radiation of the final- and initial-state particles, respectively; the latter also contain the usual parton distribution functions (PDFs).

We provided all ingredients of the factorization formula for $\tT$ at one-loop accuracy in Ref.~\cite{Becher:2015gsa}. However, to achieve N$^2$LL accuracy, one also needs their two-loop anomalous dimensions and the two-loop result for the collinear anomaly. In the present Letter, we determine these ingredients and achieve, for the first time, N$^2$LL accuracy for a transverse event shape. By fully exploiting universality properties of Eq.~(\ref{eq:fac_cs}), we manage to extract the missing ingredients from simple numerical computations. Since the same properties hold for any observable $e_\perp$, our method can be used to obtain N$^2$LL accuracy for other hadron-collider observables. Combined with numerical one-loop computations of the relevant jet, soft and beam functions, one could thus obtain an automated effective-field-theory based N$^2$LL resummation framework for hadron-collider event-shapes. Our results therefore open the door to many new studies, and several interesting applications are envisaged, as will be discussed at the end.

For the following discussion, it is useful to take the Laplace transform of the cross section, which factorizes it into a simple product rather than a convolution. Additionally, it will be crucial to consider, apart from dijet production in hadronic collisions as in Eq.~(\ref{eq:fac_cs}), also leptonic collisions, $e^+e^-\to \textrm{dijet}$, and dilepton production in hadronic collisions, $pp\to e^+e^-$. For a given partonic channel, the Laplace-transformed cross sections $\widetilde{t}(\kappa)$ for the three cases read
\begin{eqnarray}
ab\to ij\!: &\, \widetilde{t}(κ)\sim & H_{IJ}^{ab\to ij}\left(\frac{Q^2}{κ^2}\right)^{-F^{ab\to ij}(κ)}\widetilde{S}_{JI}^{ab\to ij}(κ)\nonumber\\
&&\times\widetilde{B}_a(κ)\widetilde{B}_b(κ)\widetilde{J}_i(κ)\widetilde{J}_j(κ),\label{eq:tab12}\\
e^+e^-\to ij\!: &\, \widetilde{t}(κ)\sim & H^{ij}\left(\frac{Q^2}{κ^2}\right)^{-F^{ij}(κ)}\widetilde{S}^{ij}(κ)  \widetilde{J}_i(κ)\widetilde{J}_j(κ),\nonumber\\
ab\to e^+e^-\!: &\, \widetilde{t}(κ)\sim & H^{ab}\left(\frac{Q^2}{κ^2}\right)^{-F^{ab}(κ)}\!\widetilde{S}^{ab}(κ) \widetilde{B}_a(κ)\widetilde{B}_b(κ),\nonumber
\end{eqnarray}
where the tilde always denotes a Laplace transform. We have explicitly indicated the partons upon which each of the elements in the formula depend. The hard function $H^{ab\to ij}$ does not depend on the observable and is obtained directly from the QCD amplitudes for the given partonic channel. The one-loop results of these, needed for N$^2$LL resummation, are well known. The $F^{ab\to ij}$ term is the collinear anomaly and, together with the rest of the functions, does depend on the observable. Factorization constraints require that $F^{ab\to ij}=F^{ab}+F^{ij}$ and that different partonic channels for each of the two terms are related by a global factor involving only Casimir operators~\cite{Becher:2015gsa}. Not all observables suffer from $F^{ab}$ and $F^{ij}$ anomalies. For instance for transverse thrust only $F^{ab}$ is present, while thrust minor has both. If an observable involves a final-state anomaly $F^{ij}$, it can be affected by soft recoil effects. In this case $F^{ij}$ (as well as the jet and soft functions) will depend on the recoil momentum, which would complicate the numerical procedure proposed below. For the moment, we restrict ourselves to observables which are recoil insensitive, like those defined using the broadening axis~\cite{Larkoski:2014uqa}.

The cross sections $\widetilde{t}(κ)$ must be renormalization group (RG) invariant, i.e. independent of the renormalization scale μ. Moreover, since the jet and beam functions renormalize multiplicatively, the different partonic channels are separately RG invariant. The RG equations for the various elements of the factorization formulas are all of the form
\begin{equation}\label{eq:RGgeneric}
\frac{d}{d\lnμ}\,\widetilde{f}\!\left(L,\mu\right)  =  \left[- C_f\, \gamma_{\rm cusp} L +\gamma_{f}\right] \widetilde{f}\!\left(L,\mu\right) \,,
\end{equation}
where $L:=\ln\frac{\Lambda_f}{\mu}$, $\Lambda_f$ is the characteristic scale of the function $\widetilde{f}$, $\gamma_{f}$ its anomalous dimension, and $C_f$ a combination of Casimir operators. $γ_{\rm cusp}$ is the universal cusp anomalous dimension. For those functions which are matrix valued, the above equation holds after diagonalization. By solving the RG equations we obtain the resummed cross section.
To achieve N$^2$LL resummation accuracy we need $γ_{\rm cusp}$ at three loops, $\gamma_{f}$ at two loops, the one-loop expressions for the soft, jet, and beam functions, and
the anomaly exponent at two loops. The three-loop $γ_{\rm cusp}$ is known~\cite{Moch:2004pa}. In Ref.~\cite{Becher:2015gsa} we computed the soft, jet, and beam functions for $\tT$ at one loop. In the following we describe a general method to obtain the two-loop $γ_f$'s and anomaly exponent.

Due to RG invariance, the anomalous dimensions of the different objects in Eqs.~(\ref{eq:tab12}) must sum up to zero,
\begin{eqnarray}
γ_{H^{ab\to ij}}+γ_{S^{ab\to ij}}+γ_{B_a}+γ_{B_b}+γ_{J_i}+γ_{J_j} & = & 0,\label{eq:anrelabee}\\
γ_{H^{ij}}+γ_{S^{ij}}+γ_{J_i}+γ_{J_j} & = & 0,\label{eq:anrelee}\\
γ_{H^{ab}}+γ_{S^{ab}}+γ_{B_a}+γ_{B_b} & = & 0.\label{eq:anrelpp}
\end{eqnarray}
The hard anomalous dimensions are all known at the required accuracy from the general results of Refs.~\cite{Becher:2009cu,Becher:2009qa}. To determine the remaining anomalous dimensions, we first consider the process $pp\to e^+e^-$, which is mediated by the partonic channel $q\bar{q}\toγ^*\to e^+e^-$. If we use the standard form of the analytic phase-space regulator and regularize the phase space according to~\cite{Becher:2011dz},
\begin{equation}\label{eq:regulator}
\int \!d^dk \,  \delta(k^2) \,\theta(k^0) \to \int \!d^dk \,  \delta(k^2) \,\theta(k^0) \,  \left(\frac{\nu}{n_b\cdot k}\right)^{\alpha}\,,
\end{equation}
where $n_b$ is a light-like vector in the direction of parton $b$, then the corrections to the soft function only involve scaleless integrals, since $e_{\perp}$ only depends on $\vec{k}_{\perp}$. The soft anomalous dimension $γ_{S^{ab}}$ is therefore zero. Eq.~(\ref{eq:anrelpp}) then implies that $\gamma_{B_q}$ is just given by the hard anomalous dimension. By considering as well $pp\to γγ$, mediated by the partonic channel $gg\to H\toγγ$ ($H$ represents the Higgs here), the same conclusion applies for the gluon beam function. Therefore we have $γ_{B_a}=γ^a$, where $γ^a$ are the well-known quark or gluon anomalous dimensions~\cite{Becher:2009qa}. 

Next, we consider the lepton-collider case. For observables that involve an anomaly $F^{ij}$, the soft and jet functions have the same characteristic scales. One then only needs their product, whose anomalous dimension is directly given by $γ_H$, though one will need to determine $F^{ij}$ as will be discussed below. Let us therefore consider $F^{ij}=0$, as is the case for transverse thrust. From the process $e^+e^-\to q\bar{q}$, the two-loop anomalous dimension $γ_{S^{qq}}$ can be numerically obtained by comparing the prediction of the factorization formula with the outcome of a fixed-order code like \texttt{EVENT2}~\cite{Catani:1996vz},\footnote{In the future this step can be replaced by a numerical evaluation of the two-loop soft function~\cite{SCETtalk}.} as was done for $\tT$ in Ref.~\cite{Becher:2015gsa}. The relation Eq.~(\ref{eq:anrelee}) then determines $γ_{J_q}$ at two loops. To also obtain $γ_{J_g}$, we can consider the process $e^+e^-\to H \to gg$. At two loops the soft anomalous dimension for this process can be obtained from $γ_{S^{qq}}$ by Casimir scaling, i.e. by multiplying it by the ratio of the Casimirs in the adjoint and fundamental representations, $C_A/C_F$, since the only difference between the two cases is the color representation of the soft Wilson lines. Equation~(\ref{eq:anrelee}) then determines $γ_{J_g}$. Explicitly for $\tT$~\cite{Becher:2015gsa},
\begin{equation}
γ_{1 J_g}=γ_1^g-C_A^2\left(74^{+15}_{-10}\right)+C_AT_Fn_f\left(9^{+1.5}_{-1.0}\right),
\end{equation}
where $T_F=1/2$, $n_f$ is the number or light flavors, and the anomalous dimensions are expanded in the strong coupling $α_s$ as $γ=\sum_nγ_n(α_s/4π)^{n+1}$. With the anomalous dimensions of the hard, jet and beam functions at hand, we then immediately obtain the soft-function anomalous dimensions for any partonic channel $ab \to ij$ from Eq.~(\ref{eq:anrelabee}).

\begin{figure*}
\includegraphics[height=5.5cm]{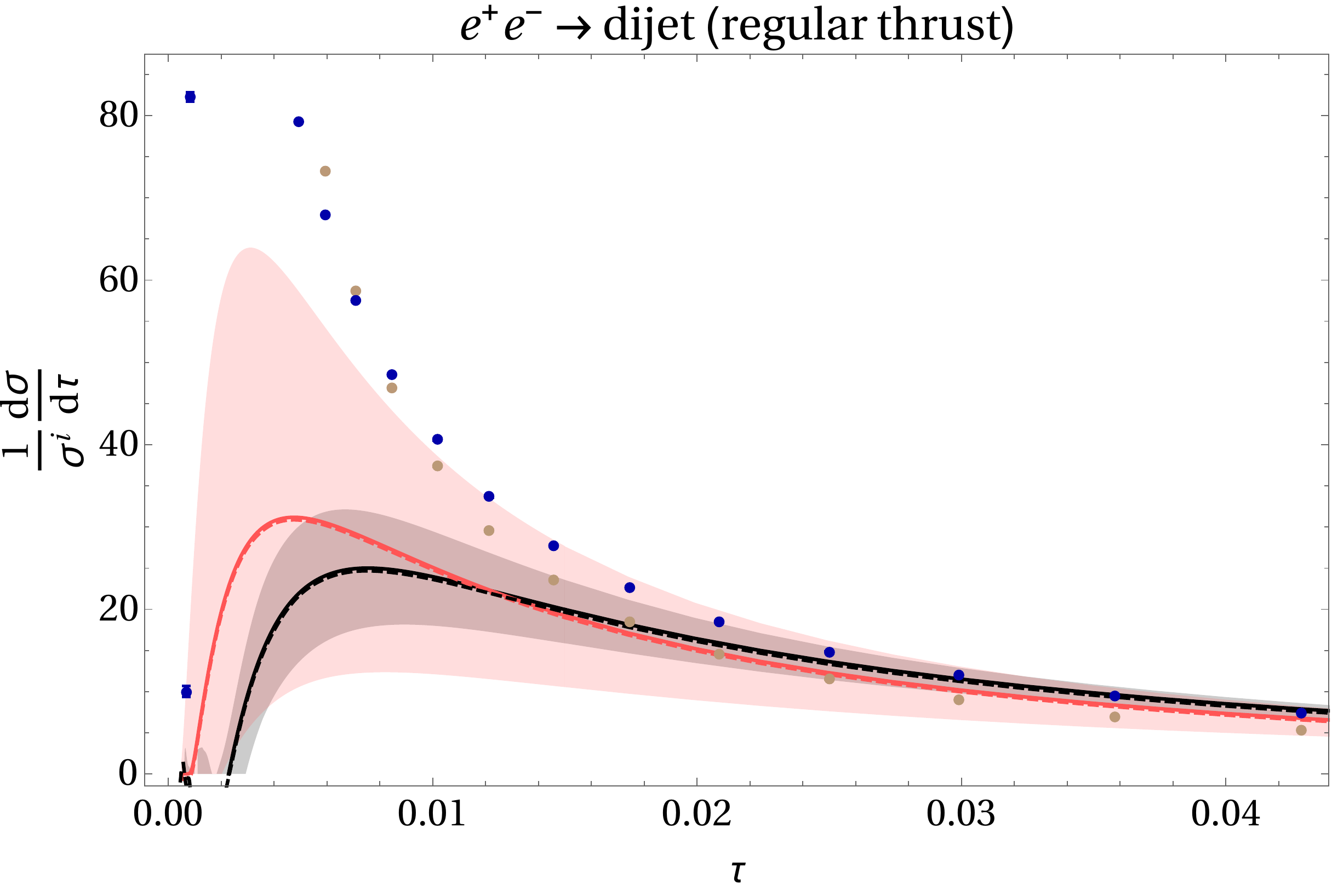}
\includegraphics[height=5.5cm]{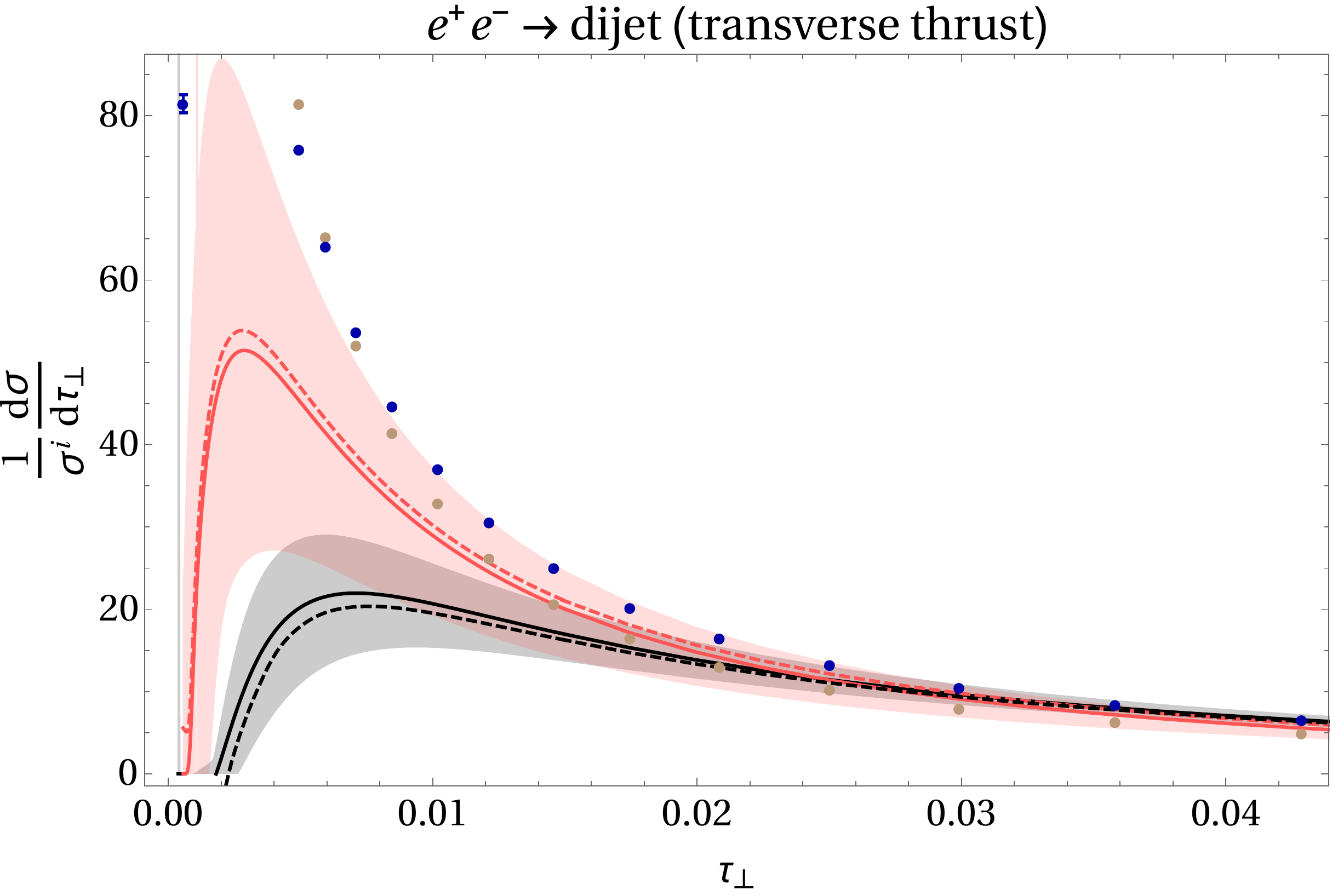}
\caption{Resummed cross section for $e^+e^-\to$~dijet at NLL accuracy (red/lighter lines and bands) and at N$^2$LL accuracy (black/darker lines and bands), together with the fixed-order results at $\mathcal{O}(α_s)$ (brown/lighter points) and at $\mathcal{O}(α_s^2)$ (blue/darker points). Left panel: regular thrust; right panel: $\tT$. The solid lines are always pure resummation, and the dashed ones (barely visible in the regular-thrust case) are the resummed results with additive matching to the fixed-order calculation obtained with \texttt{EVENT2}.}\label{fig:rescs_lin}
\end{figure*}

The last missing ingredient for N$^2$LL accuracy is the two-loop anomaly. We explain in the following how to obtain it for transverse thrust using the process $pp\to e^+e^-$. For recoil-free observables that also involve a final-state anomaly $F^{ij}$ an analogous procedure is repeated in the leptonic case. (For recoil-sensitive observables the method would need to be repeated for different recoil values to reconstruct the anomaly as a function of recoil.) 
The strategy to obtain $F^{ab}$ is to define a new observable that coincides with $\tT$ for one emission and for which the anomaly is known, and then compute the difference of this new observable with $\tT$. Since the computation of this difference only involves multiple emission diagrams, it reduces at order $α_s^2$ to evaluating a double-emission tree-level contribution. The same strategy was used in Refs.~\cite{Banfi:2012yh,Banfi:2012jm,Becher:2013xia}, in the context of jet-veto cross sections. Denoting the momentum of the lepton pair by $q$, we define a new observable $\mathcal{S}_{\perp}$, according to
\begin{equation}
\mathcal{S}_{\perp}:=|\vec{q}_{\perp}|-|\vec{q}_{\perp}\cdot\vec{n}_{\perp}|=|\sum_m\vec{p}_{m\perp}|-|\sum_m\vec{p}_{m\perp}\cdot\vec{n}_{\perp}|,
\end{equation}
where the sums run over hadronic final states. For transverse thrust we instead have
\begin{equation}\label{eq:taucal}
\mathcal{T}_{\perp}:=Q_{\perp}\tT=\sum_m\left(|\vec{p}_{m\perp}|-|\vec{p}_{m\perp}\cdot\vec{n}_{\perp}|\right),
\end{equation}
where, for  $\tT\to0$, $Q_{\perp}=E_{e^+}|\sin\theta_{e^+}|+E_{e^-}|\sin\theta_{e^-}|$, with $E_{e^{\pm}}$ and $\theta_{e^{\pm}}$ the energies and angles with respect to the beam of the two leptons. Note that the electron and the positron define the thrust axis, i.e.\ at leading order in the SCET power counting they are the only particles with large transverse momentum for $\tT\to0$, and therefore do not need to be included in the sum over $m$ in Eq.~(\ref{eq:taucal}). It is then clear that $\mathcal{S}_{\perp}$ and $\mathcal{T}_{\perp}$ start differing when there are at least two emissions. Since $\mathcal{S}_{\perp}$ is defined in terms of $\vec{q}_{\perp}$, it can be obtained at two-loop accuracy from the known results for the Drell-Yan cross section $σ^{DY}$ at small transverse momentum, which in Laplace space has a factorized form with a structure analogous to Eq.~(\ref{eq:tab12}).
We can translate $σ^{DY}$ into a result for $\mathcal{S}_{\perp}$ by inserting a δ function enforcing the appropriate constraint and integrating over $\vec{q}_{\perp}$
\begin{equation}\label{eq:SperpfrDY}
\frac{dσ}{d\mathcal{S}_{\perp}}\!=\!\!\!\int\! d^2q_{\perp}δ(\mathcal{S}_{\perp}-|\vec{q}_{\perp}|(1-|\cos\phi|))\frac{d^2σ^{DY}}{d^2q_{\perp}},
\end{equation}
where $\phi$ is the angle between $\vec{q}_{\perp}$ and $\vec{n}_{\perp}$. From Eq.~(\ref{eq:SperpfrDY}) we obtain the fixed-order expansion of the $\mathcal{S}_{\perp}$ distribution at two loops. This expression involves the two-loop anomaly exponent 
\begin{equation}
F^{q\bar{q}}(L)=\frac{α_s}{4π}C_FΓ_0L+\left(\frac{α_s}{4π}\right)^2\!C_F\!\left(Γ_0β_0\frac{L^2}{2}+Γ_1L+d_2^q\right),
\end{equation}
which we have written in terms of a logarithm $L$ of the relevant variable for the process. In this expression $Γ_i$, $β_i$ are the coefficients of $γ_{\rm cusp}$ and the beta function, and $d_2^q$ was determined in Ref.~\cite{Becher:2010tm}. An analogous expression can be obtained for the expansion of the $\mathcal{T}_{\perp}$ distribution. It has the same structure but involves an unknown two-loop constant $d_2^{\perp}$, which we want to determine. With these expressions we can compute the difference of the two distributions. Evaluating it explicitly, we verify that it vanishes at the one-loop level. At the two-loop level it involves a term proportional to $d_2^q-d_2^{\perp}$ which is accompanied by a logarithm of $Q$. On the other hand, we can also compute this logarithmic piece of the difference directly. The logarithm is due to the rapidity difference between soft and collinear emissions, and its coefficient can be extracted from the rapidity divergences of the beam or soft functions. It is convenient to extract it from the soft function, since the relevant amplitudes are simpler. With the standard form of the analytic regulator, Eq.~(\ref{eq:regulator}), the soft function is scaleless and vanishes, but we can use a  symmetric regulator
\begin{equation}\label{eq:splitreg}
\int \!d^dk\, \delta(k^2)\theta(k^0)\! \left[    \left( \frac{\nu}{ n_b \cdot k } \right)^\alpha\!\!\! \theta(n_b \cdot k - n_a \cdot k) + (a\leftrightarrow b) \right],
\end{equation}
for which the soft function is nonzero and its divergences cancel against the ones of the beam functions. To perform the calculation, we need the tree-level two-emission soft amplitude squared, which was given in a convenient form in Refs.~\cite{Becher:2012qc,Tackmann:2012bt}. It is easy to isolate and perform the integration that produces the $1/α$ divergence in the analytic regulator, and the remaining integrations can be performed numerically (we used the Cuba library~\cite{Hahn:2004fe}). Combining these results with the known value of $d_2^q$~\cite{Becher:2010tm}, we obtain\footnote{We thank Guido Bell and Rudi Rahn for crosschecking this result and pointing out an algebraic mistake, which affected the value in the original version of this paper.} the two-loop anomaly coefficient for $\tT$
\begin{equation}\label{eq:d2perp}
d_2^{\perp}=(208.0\pm0.1)\,C_A+(-37.191\pm0.006)\,T_Fn_f.
\end{equation}
We have cross-checked this result by comparing the prediction of the factorized formula with the output of the fixed-order code DYNNLO~\cite{Catani:2009sm,Catani:2007vq} at low values of $\tT$ and find good agreement. 

Having obtained all the two-loop anomalous dimensions and the two-loop anomaly exponent, we have established N$^2$LL resummation accuracy for $\tT$ at hadron colliders. It is worth emphasizing the huge simplifications that separating the effects from the different relevant physical scales brought about. After computing the one-loop soft, jet, and beam functions, this allowed us to obtain all the required two-loop coefficients using a fixed-order code for leptonic collisions, with no need to perform new two-loop calculations.

\begin{figure*}
\includegraphics[height=5.5cm]{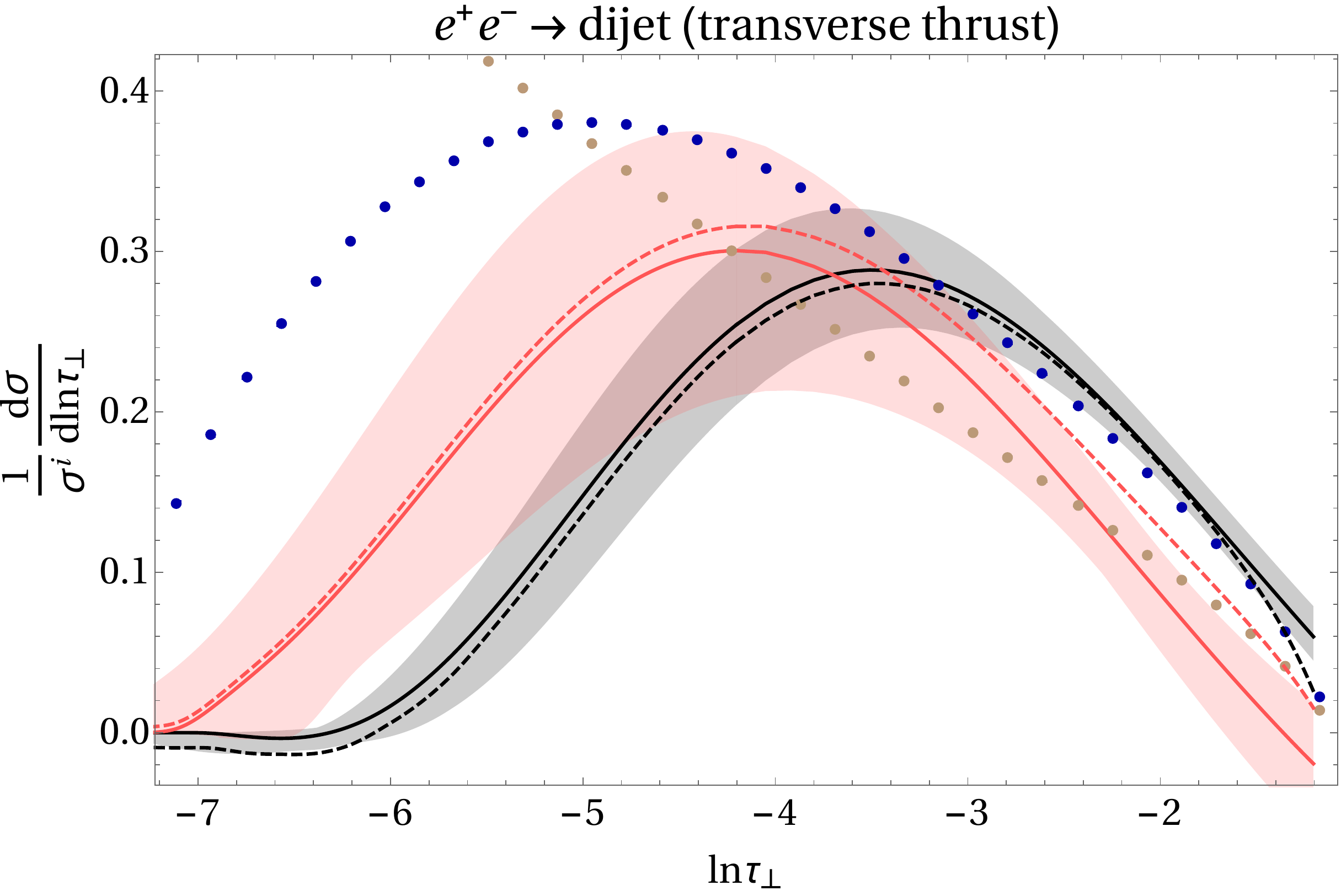}
\includegraphics[height=5.5cm]{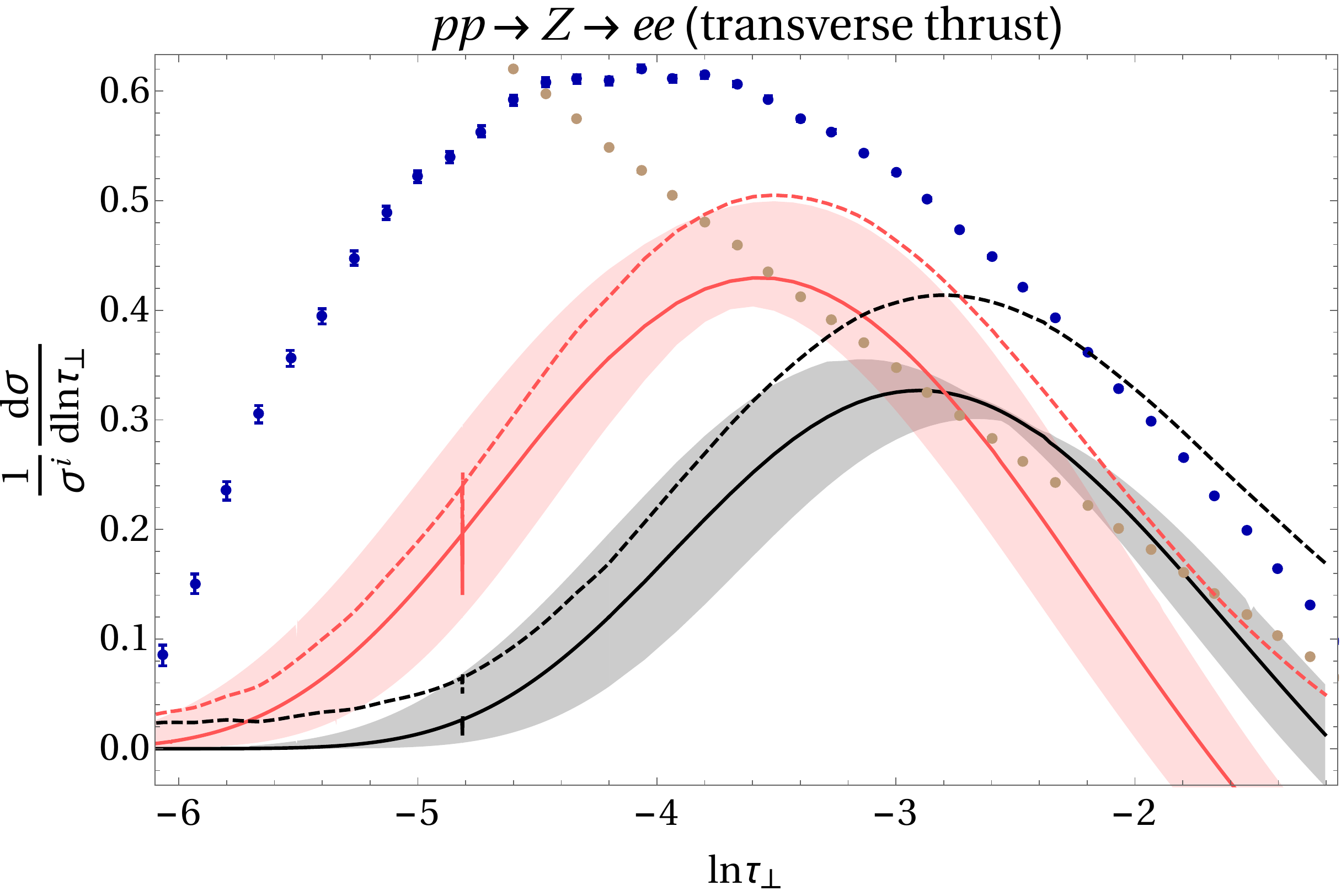}
\caption{Transverse thrust cross section for $e^+e^-\to$~dijet (left) and $pp\to Z\to e^+e^-$ (right), color coding as in Fig.~\ref{fig:rescs_lin}. The fixed-order results are computed with DYNNLO.}\label{fig:rescs_log}
\end{figure*}

Let us now illustrate the effect of the resummation for transverse thrust. Since there are several ingredients that enter in the factorized formulas, it is instructive to consider: (1) $e^+e^-\to q\bar{q}$, which involves only soft and jet functions, and (2) $pp\to e^+e^-$, which complementarily involves only beam functions and the anomaly. We show in Fig.~\ref{fig:rescs_lin} the resummed cross section in the lepton-collider case, compared with the corresponding results for regular thrust. Fig.~\ref{fig:rescs_log} shows transverse thrust, for both the $e^+e^-$ and $pp$ case, plotted as a function of $\ln \tT$.
The plots are obtained by taking the derivative of the resummed expression for the cumulant, and we normalize the N$^{i+1}$LL result to $σ^i$, the total cross section including $\mathcal{O}(α_s^i)$ corrections. Our default scale choices are $\mu_{\rm soft}=4Q\tT$, $\mu_{\rm jet}=2Q\sqrt{\tT}$, in the $e^+e^-$ case, and $\mu_{\rm beam}=2c_0Q\tT$ in the $pp$ case ($\ln c_0:=4G/\pi$, where $G$ is the Catalan constant); they are set after integration over the scattering angle, i.e. the angle of the thrust axis to the beam axis (see~\cite{Becher:2015gsa} for more details on the default scale choices). We set the hard scale to $Q=M_Z$, use NNLO MSTW 2008 PDFs~\cite{Martin:2009iq}, and $α_s(M_Z)=0.11707$. The bands in the plots represent the changes induced by varying each of the scales by a factor of two and then adding the individual scale-variation bands in quadrature.\footnote{We work with $n_f=5$ but for small $\tT$ some scales cross the $b$-quark threshold, which leads to small spikes in the plots. To avoid this unphysical behavior, one should decouple heavy flavors, as was done for non-anomalous processes e.g. in Ref.~\cite{Gritschacher:2013pha}. We leave this for future work.} 
Let us note that in the N$^2$LL $pp$ curve we exponentiated the $\mathcal{O}(α_s)$ terms in the anomaly and the $δ$-function part of the beam functions. Without this exponentiation, the large one-loop corrections would lead to a negative cross section below $\ln\tT\approx-4.5$. Quite generally, we find that the N$^2$LL corrections are large, especially in the hadron-collider case and that the lower-order scale-uncertainty bands tend to underestimate these effects. For completeness, we also included results matched to the fixed-order computation, N$^i$LL+$\mathcal{O}(α_s^i)$ in the $e^+e^-$ case and N$^i$LL+$\mathcal{O}(α_s)$ in the $pp$ case; if future analyses require it, the matching could be extended to $\mathcal{O}(α_s^3)$ in the $e^+e^-$ case using \texttt{EERAD3}~\cite{Ridder:2014wza} and at $\mathcal{O}(α_s^2)$ in the $pp$ case using DYNNLO. We can clearly see that resummation is a quite important effect in both $e^+e^-$ and $pp$ cases. A comprehensive phenomenological analysis of the dijet case and other phenomenologically relevant processes like $pp\to Z+\textrm{jet}$~\cite{Chatrchyan:2013tna} or $pp\to ZZ$, is beyond the scope of this Letter and will be presented elsewhere. All the ingredients needed to produce the resummed results for the different processes are being implemented into a numerical code, to be made publicly available in the future. 

The factorization theorem Eq.~(\ref{eq:fac_cs}) resums all large logarithms in the partonic cross section but assumes that the hadronic cross section is obtained by convoluting with PDFs. It has been argued that for hadronic event shapes also Glauber gluons will contribute~\cite{Gaunt:2014ska,Zeng:2015iba}, which are not captured by the standard factorization. The corresponding effects should be analyzed in SCET and included if present. A comparison of our results with data may shed some light on the issue of Glauber-gluon effects and help to clarify their relation to UE effects. A better understanding of these effects should also help to assess to what extent UE effects are mitigated when certain combinations of event shapes are used~\cite{Aaltonen:2011et}. For this purpose, $\tT$ in $pp\to e^+e^-$ shown here is particularly useful, since it is one of the simplest processes that is affected by UE effects. It would thus be interesting to have precise LHC data for it and to carefully compare with our results. Once the UE is better understood, one could construct a combination of event shapes that is as insensitive to it as possible to obtain a novel determination of $α_s$ at much higher energies than what has been done before with leptonic event shapes. 

\begin{acknowledgments}
We thank the high-energy theory group at Harvard and the Mainz Institute for Theoretical Physics (MITP) for hospitality and support. This work is supported by the Swiss National Science Foundation (SNF) under the Sinergia grant number CRSII2\_141847\_1.
\end{acknowledgments}

\end{document}